# Stability of geometrically frustrated magnetic state of $Ca_3CoRhO_6$ to applications of positive and negative pressure

Niharika Mohapatra, Kartik K Iyer, Sudhindra Rayaprol,[1] S. Narayana Jammalamadaka, and E.V. Sampathkumaran[*]

*Tata Institute of Fundamental Research, Homi Bhabha Road, Colaba, Mumbai – 400005, India*
*[1]UGC-DAE Centre for Scientific Research, BARC Campus, Trombay, Mumbai – 400085, India*

The influence of negative chemical pressure induced by gradual replacement of Ca by Sr as well as of external pressure (up to 10 kbar) on the magnetism of $Ca_3CoRhO_6$ has been investigated by magnetization studies. It is found that the solid solution, $Ca_{3-x}Sr_xCoRhO_6$, exists at least till about $x= 1.0$ without any change in the crystal structure. Apart from insensitivity of the spin-chain feature to volume expansion, the characteristic features of geometrical frustration interestingly appear at the same temperatures for all compositions, in sharp contrast to the response to Y substitution for Ca (reported previously). Interestingly, huge frequency dependence of ac susceptibility known for the parent compound persists for all compositions. We do not find a change in the properties under external pressure. The stability of the magnetic anomalies of this compound to the volume change (about 4%) is puzzling.

# I   INTRODUCTION

The compounds crystallizing in the $K_4CdCl_6$-type rhombohedral structures have started attracting considerable attention in recent years, as, in this family, one observes several manifestations of interplay between spin-chain behavior and topologically frustrated magnetism due to triangular arrangement of (antiferromagnetically-coupled) magnetic chains [1]. Among these, the Co-based compounds, $Ca_3CoRhO_6$ and (to a larger extent) $Ca_3Co_2O_6$, have been of special interest to the community due to their interesting magnetic behavior [2, 3]. As the temperature is lowered, following short-range correlation features over a wide temperature range arising from the spin-chains, one observes two magnetic transitions (which we label as $T_1$ and $T_2$). At $T_1$, two of the three chains at the corners of the triangle undergo magnetic ordering with an intrachain ferromagnetic interaction and interchain antiferromagnetic coupling, while the third one is left incoherent, leading to a not-so-common 'partially disordered antiferromagnetic (PDA) structure' (see articles cited in Refs. 3 and 4 for few more examples for PDA structure, $CsCoCl_3$ and $CsCoCl_3$). At a lower temperature, $T_2$, a complex glassy phase evolves which results in an unusually large frequency dependence of ac susceptibility ($\chi$), revealing interesting spin-dynamics [4,5]. Though the values of $T_1$ and $T_2$ are widely different for these two compounds (near 24 and 10 K for $Ca_3Co_2O_6$ and 90 and 30 K for $Ca_3CoRhO_6$), the dc $\chi$ curves, if plotted as a function $T/T_1$, are found to be similar [6]. While large values of $T_1$ and $T_2$ for the latter could be explained through LSDA calculations including Coulomb interaction and spin-orbit coupling [7], this similarity among these compounds is quite striking considering that, in the former, one of the two Co sites (running alternately along $c$-axis) is non-magnetic, whereas, for the latter, the consensus is that there is a magnetic moment at both Co and Rh sites ($Co^{2+}$, $3d^7$ high-spin; $Rh^{4+}$, $4d^5$ low spin) [7-9].

In this article, we investigate whether the behavior of $Ca_3CoRhO_6$ can be modified by chemical and external pressure effects. It is commonly known that the anomalies associated with geometrically frustrated magnetic state is quite sensitive to any perturbation or lattice distortion, and that one can even attain (long-range magnetic) order by (chemical) disorder. Even for this family, order-by-disorder has been demonstrated, for example, for $Ca_3Co_{1-x}Mn_xO_6$ (Ref. 10) and $Ca_3CuMnO_6$ (Ref. 11). In fact, even in the case of $Ca_3CoRhO_6$, a small substitution (less than 5 atomic perfect) of Y for Ca in $Ca_3CoRhO_6$ was reported to dramatically depress the high-temperature three-dimensional ordering (occurring at $T_1$) [12]. In view of these, we undertake this investigation. To simulate lattice expansion, we succeeded in gradual replacement of Ca by Sr resulting in the formation of the solid solution, $Ca_{3-x}Sr_xCoRhO_6$, without a change in the crystal structure till $x= 1.0$. For contraction, we applied external pressure up to 10 kbar. It is interesting to note that all the characteristic features of the parent compound are retained for all compositions and pressures nearly at the same temperatures as those in the parent compound - in sharp contrast to the observation in Y substituted $Ca_3CoRhO_6$. This reveals that the magnetic properties are quite robust to volume changes, but possibly sensitive to electron doping.

## II.   EXPERIMENTAL DETAILS

Attempts were made to synthesize several compositions of the series, $Ca_{3-x}Sr_xCoRhO_6$ ($x= 0.0, 0.1, 0.2, 0.3, 0.5, 1.0, 1.5, 2.0, 2.5$ and $3.0$) by a conventional solid state route as described in Ref. 4 starting from stoichiometric amounts of high purity (>99.9%) $CaCO_3$, $SrCO_3$, CoO and Rh powder. Characterization of the materials was carried out by x-ray diffraction (Cu $K_\alpha$) and scanning electron microscope (SEM). It is found that the solid solution in the rhombohedral structure forms till $x= 1.0$ only, without any other phase, whereas additional weak lines in the x-ray diffraction pattern are seen for higher compositions. For this reason, we restrict our discussions for $x \leq 1.0$. Energy dispersive x-ray analysis on several grains confirms homogeneity of the compositions. The dc magnetic susceptibility measurements in the temperature interval 1.8-300 K were performed employing a commercial superconducting quantum interference device (SQUID) (Quantum Design, USA) for zero-field-cooled (ZFC) and field-cooled (FC) conditions of the specimens for two externally applied magnetic fields ($H=$ ) 100 Oe and 5 kOe. Frequency ($\nu$) dependence of ac $\chi$ was obtained in the temperature region of interest with the same SQUID magnetometer. We have also performed isothermal M measurements at 5 and 62 K with a vibrating sample magnetometer (VSM, Oxford Instruments, U.K.), which enabled us to take the data up to 120 kOe. The magnetization behavior under high pressure for the parent compound was tracked (1.8 – 300 K) in a hydrostatic pressure medium (daphne oil) up to 10 kbar employing a pressure cell procured from EasyLab Technologies Ltd (U.K) with the help of the SQUID magnetometer. The pressure monitored by measuring superconducting transition temperature of Sn is the one being mentioned in this paper (At 300 K, the actual pressure is higher by about 2 kbar).

### III. RESULTS AND DISCUSSION

The x-ray diffraction patterns are shown in figure 1 for selected compositions ($x \leq 1.0$). It is clear from this figure that all the diffraction lines could be indexed to $K_4CdCl_6$-type rhombohedral structure. The fact that there is a gradual expansion of the unit-cell with increasing Sr concentration is convincingly demonstrated by showing the data in the higher angle side in an expanded form. This plot also demonstrates that there is a complete formation of the solid solution, as the diffraction pattern for any composition is not a superimposition of the patterns of any other compositions. The lattice constants obtained by least-squares fitting of the lines are included in the figure. Both the cell parameters, $a$, $c$, and unit-cell volume ($V$) increase with $x$, by about 1.58%, 0.69% and 4.06% respectively as $x$ is varied from 0 to 1.

The results of dc $\chi$ measurements are shown in figure 2 in two different ways. For all compositions, we see all the features noted for the parent compound [3]: There is a Curie-Weiss region above 225 K, followed by a broad peak in $\chi(T)$ in the range 100 to 150 K representative of intrachain interaction strength [13]; there is a sudden increase of $\chi$ below 100 K attributable to the onset of three dimensional ordering [3], followed by a peak in ZFC $\chi(T)$ curve at 40 K with a bifurcation of ZFC-FC curves as known earlier (see, for example, Ref. 4) due to the onset of second magnetic transition of a glassy type. In the range 30 to 90 K, M/H is H-dependent known for PDA structure. A visual inspection of the curves itself clearly reveals that the features do not change with Sr substitution. In order to demonstrate more vividly that the $T_1$ and $T_2$ do not vary across the solid solution, we have plotted $d\chi/dT$ of $H= 100$ Oe data as well in figure 2. It is clear that the plots for all compositions overlap and, in particular, there is a sudden change in the value of $d\chi/dT$ at the same temperatures. This establishes that these transition temperatures remain the same for all $x$. (From this derivative plot, we infer that $T_1$ could be actually 100 K, and not 90 K as proposed in the literature). It is therefore evident that interchain coupling is not sensitive to Sr substitution. The sign and magnitude of the paramagnetic Curie temperature ($\theta_p = 136 \pm 2$ K ) obtained from the Curie-Weiss region (225 – 300 K) also do not vary with $x$. Since the sign of $\theta_p$ is positive, this parameter can be directly correlated to intra-chain ferromagnetic exchange coupling in these oxides. Incidentally, the magnitude of $\theta_p$ falls in a temperature range where $\chi(T)$ exhibits a maximum. Taking these observations into consideration, we conclude that the intrachain coupling strengths also are unaffected by lattice expansion. Finally, the magnitude of the effective moment (5.3 ± 0.1 $\mu_B$ per formula unit) obtained from high temperature Curie-Weiss region is the same for all the compositions within the limits of experimental error, which establishes that the spin/valence states of Co and Rh do not change with $x$.

We address whether the 'superparamagnetic-like' ac $\chi$ response [4] is modified across the series. The real ($\chi'$) and imaginary ($\chi''$) parts of ac $\chi$ as a function of temperature are shown in figure 3 for $H = 0$. As known earlier [4], for the parent compound, there is a peak in $\chi'$ and $\chi''$ at about 50 and 38 K respectively for $\nu= 1.3$ Hz, and these peak temperatures undergo a dramatic upward shift (by about 12 K) as $\nu$ is increased to 1.333 kHz. These features interestingly persist for all compositions, as though the spin-dynamics also is not influenced at all by lattice expansion. Since the peak in $\chi''$ is rather sharp compared to that in $\chi'$, it is more straightforward to infer the trends in the $x$-independence of characteristic temperatures from the peak position in $\chi''$. We have also performed ac $\chi$ measurements in the presence of high magnetic fields (not shown here) and the features are not different from that noted for $x= 0.0$ (Ref. 4).

In order to investigate whether positive pressure induces a change in the magnetic behavior, we have performed high pressure dc magnetization studies at 2, 6 and 10 kbar. The observation of emphasis is that the features are unaffected by compression of the lattice caused by the external pressure as well (see figure 4). The fact that there is no observable variation in $T_1$ and $T_2$ is established by the fact the curves of $d\chi/dT$ lie one over the other for all pressures, as demonstrated in an inset of figure 4.

In this connection, it is worthwhile recalling that Martinez et al [14] investigated influence of high pressure on the magnetic anomalies of $Ca_3Co_2O_6$. It was found that there is a small upward shift of magnetic ordering temperature ($T_N$) and the magnetic field, $H_{SF}$, at which spin reorientation effect occurs as inferred from the isothermal M data in the PDA regime. These authors used the values of $H_{SF}$ as a measure of interchain coupling and hence of magnetic ordering temperatures. $H_{SF}$ was assumed [15] to be related to compressibility modulus. Therefore, following Ref. 14,

$$\frac{d \ln T_N}{dP} \approx \frac{d \ln H_{sf}}{dP} = -\frac{n}{l}\frac{dl}{dP}$$

where $l$ is interchain distance and the value of $n$ is typically 12. Interestingly, a quantitative agreement between the experimentally observed variations in transition temperatures and those derived from $H_{SF}$ was found with the help of this relationship. The pressure derivative of magnetic ordering temperature was found to be 0.073 K/kbar. In the compound under investigation, $Ca_3CoRhO_6$, the value of $H_{SF}$ is about 55 kOe which is the extreme limit of the magnetic field in the high pressure experiments; hence it was not possible for us to track an increase of $H_{SF}$ with external pressure. Therefore, we used isothermal M data (Figure 5) obtained up to 120 kOe (with VSM) for the Sr-substituted oxides, which allows us to draw some conclusions with the knowledge of influence of negative pressure. Though all the 5K-plots overlap for all Sr substituted specimens (see inset of figure 5), the value of $H_{SF}$ varies wih $x$ significantly in the PDA region, say at 62 K: ~ 55, 52, 51, 45, and 40 kOe for $x=$ 0.0, 0.1, 0.3, 0.5, and 1.0 respectively. In the absence of compressibility data for this compound, we use the value employed in Ref.14 for this property. Then, the value of $c$ for -10 kbar corresponds to a composition close to $x = 1.0$. This means that, for this compound, $d\ln H_{sf}/dP = 0.027$ kbar$^{-1}$. Following above-mentioned relationship between ordering temperature and $H_{sf}$, one should have observed a depression of about 25 K for $T_1$ as $x$ varies from 0.0 to 1.0. This is not found to be the case experimentally. (A more straightforward comparison of the behavior of $Ca_3Co_2O_6$ and $Ca_3CoRhO_6$ can be made by noting that, for the former, as shown in the figure 3a of Ref. 14, $dH_{sf}/dP$ falls in the range of 135 -150 Oe/kbar, whereas in the latter, it appears to be about 10 times larger as inferred from the data for Sr samples). Clearly, this line of arguments leads us to propose that the observed insensitivity of magnetism to positive and negative pressure resulting in a volume change of at least 4% (which is of significant magnitude), is intriguing.

## IV. CONCLUSION

Positive and negative pressure, resulting in a significant unit-cell volume change of at least 4%, does not influence the features attributable to geometrical frustration and spin-chain interaction as well as the characteristic magnetic transition temperatures in the compound, $Ca_3CoRhO_6$. This is despite the fact that the temperatures governing intrachain and interchain interactions appear to be about three times larger than in $Ca_3Co_2O_6$, in which case an observable variation in the properties were reported in the literature. The extreme robustness of the properties of the former is noteworthy. A different behavior of Y-substituted systems [12] could then be attributed to sensitivity to electron doping. It is notable that possible randomness of interatomic distances or local strain induced by chemical-disorder in the Sr-based solid solution does not lift frustration effects in favor of long-range magnetic order. We emphasize that unusually large frequency dependence of the peak in ac susceptibility in the parent compound, that too without getting influenced by chemical doping, is an issue that requires urgent theoretical attention.

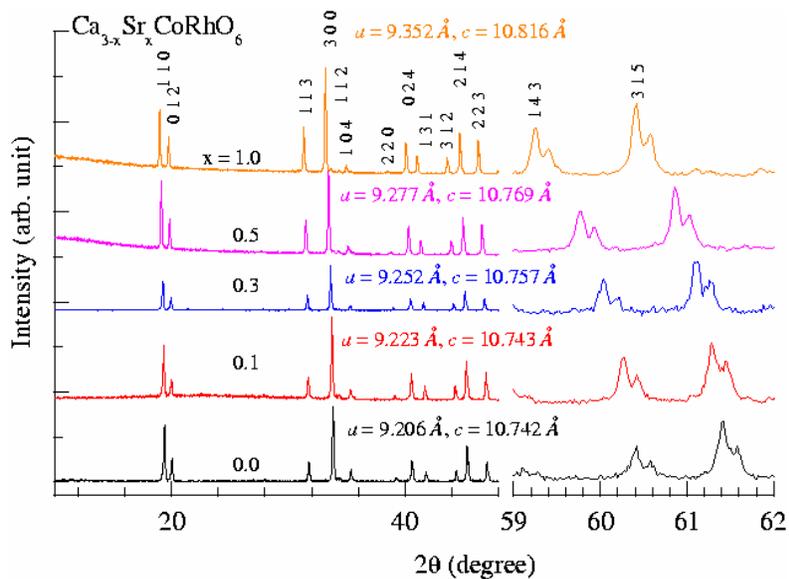

Figure 1: (color online) X-ray diffraction patterns (Cu $K_\alpha$) of solid solution, $Ca_{3-x}Sr_xCoRhO_6$ (x= 0.0, 0.1, 0.3, 0.5, and 1.0). The patterns at the higher angle side are shown in an expanded form (and the splitting of the lines is due to slightly different wavelengths of $K_{\alpha1}$ and $K_{\alpha2}$). The lattice constants ($\pm$ 0.004 Å) and the Miller indices are included.

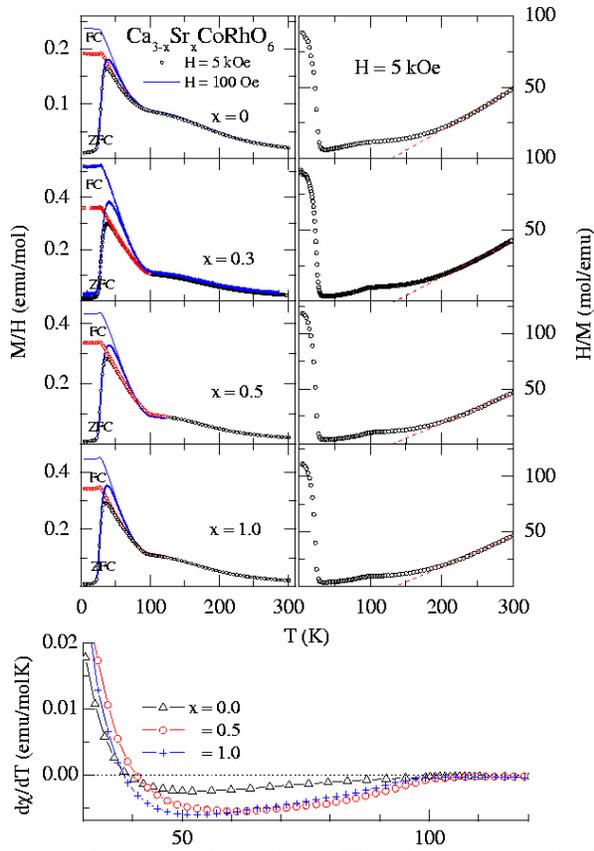

Figure 2: (color online) **(Right)** Dc magnetization (M) divided by magnetic field (H) as a function of temperature for the solid solution, $Ca_{3-x}Sr_xCoRhO_6$ ($x \leq 1.0$), for the zero-field-cooled and field-cooled conditions of the specimens, in the presence of 100 Oe and 5 kOe. **(Left)** ZFC H/M as a function of temperature. A discontinuous line is drawn through the high temperature linear region. **(Bottom)** The temperature derivative of magnetic susceptibility (ZFC, H= 100 Oe) as a function of temperature in the vicinity of two magnetic transitions for three selected compositions.

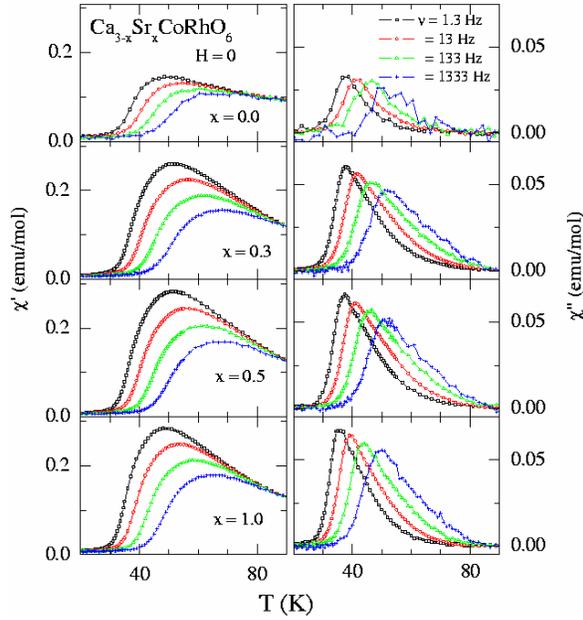

Figure 3: (color online) Real ($\chi'$) and imaginary ($\chi''$) parts of ac susceptibility for $Ca_{3-x}Sr_xCoRhO_6$ measured with different frequencies and an ac field of 1 Oe. The curves move towards higher temperatures with increasing frequency ($\nu$). The lines through the data points serve as a guide to the eyes.

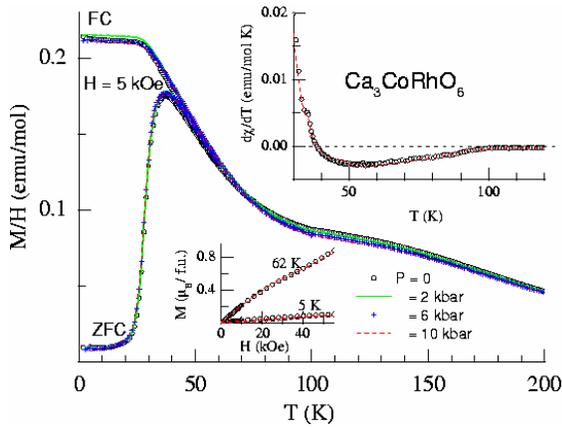

Figure 4: (color online) Dc magnetization (M) divided by magnetic field (1.8 – 200 K) under pressure for the solid solution, $Ca_3CoRhO_6$, for the zero-field-cooled and field-cooled conditions of the specimens, in the presence of 5 kOe. The data above 200 K also overlap for all pressures, and hence are not shown. In the top inset, the temperature derivatives of magnetic susceptibility in the range 30 – 120 K are plotted for zero pressure and 10 kbar. In the bottom inset, for the same pressures, isothermal magnetization data (ZFC condition) at 5 and 62 K are plotted.

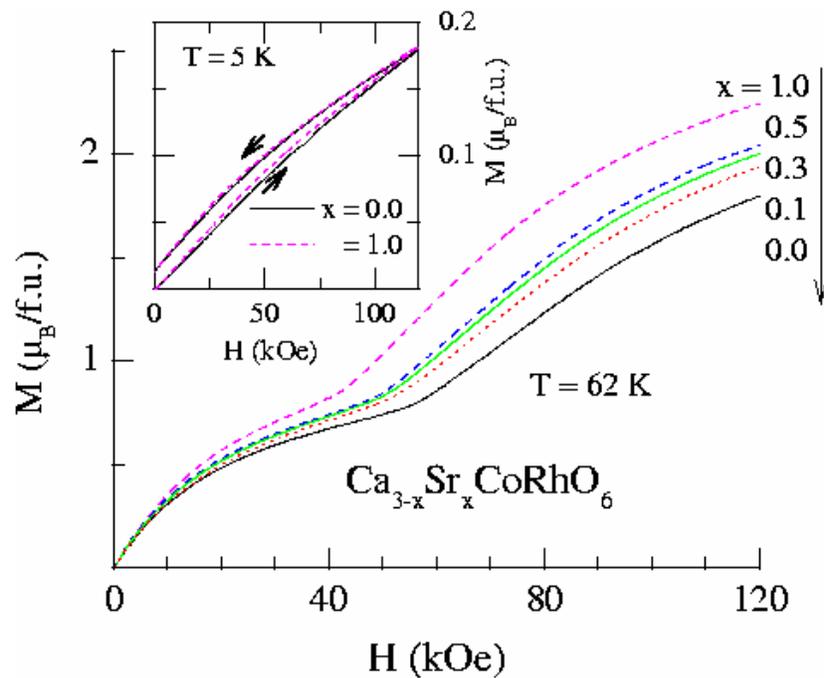

Figure 5: (color online) Isothermal magnetization at 5 (in the inset) and 62 K for Ca$_{3-x}$Sr$_x$CoRhO6 (x ≤ 1.0) for the zero-field-cooled conditions of the specimens under ambient pressure. The curves at 62 K are nonhysteretic.